\documentstyle[multicol,aps,prb]{revtex}

\newcommand{\th}{\theta}
\newcommand{\tm}{\theta_{max}}
\newcommand{\tn}{\theta_n}

\newcommand{\g}{\Gamma}
\newcommand{\hti}{\tilde{h}}
\newcommand{\R} [1] {~\mbox{(\ref{#1})}}
\newcommand{\C} [1] {~\mbox{[\onlinecite{#1}]}}
\newcommand{\vuh}{v_{uh}}

\begin{document}

%
\title{Thick surface flows of granular materials: \\
The effect of the velocity profile on the avalanche amplitude}
%
\vspace{1cm}
\author{Achod Aradian, Elie Rapha\"el and Pierre-Gilles de Gennes}
\address{Laboratoire de Physique de la Mati\`ere Condens\'ee,
URA $n^o 792$ du C.N.R.S., \\ Coll\`ege de France, 11 place
Marcelin Berthelot, 75231 Paris Cedex 05, France \\ {\it e-mail}:
aradian@ext.jussieu.fr, elie@ext.jussieu.fr,
Pierre-Gilles.DeGennes@espci.fr}
\date{\today}
\maketitle
\begin{abstract}
A few years ago, Bouchaud {\it et al.} introduced a
phenomenological model to describe surface flows of granular
materials \mbox{[J. Phys. Fr. I {\bf 4}, 1383 (1994)]}. According
 to this model, one can distinguish between a static phase and a rolling phase that are able to
exchange grains through an erosion/accretion mechanism. Boutreux
{\it et al.} \mbox{[Phys. Rev. E {\bf 58}, 4692 (1998)]} proposed a
modification of the exchange term in order to describe thicker
flows where saturation effects are present. However, these approaches assumed that the downhill
convection velocity of the grains is constant inside the rolling phase, a hypothesis that is not
verified experimentally. In this article, we therefore modify the above models by introducing a
velocity profile in the flow, and study the physical consequences of this modification in the simple
situation of an avalanche in an open cell. We present a complete analytical description
of the avalanche in the case of a linear velocity profile, and generalize the results for a
power-law dependency. We show, in particular, that the amplitude of
the avalanche is strongly affected by the velocity profile.
\par
{\bf Short title}: Thick granular surface flows
\end{abstract}


\begin{multicols}{2}
\section{General principles}
\subsection{Onset of avalanches}
\label{Onset}
\par It is a dailylife experience that the top surface of a mass of
granular matter need not be horizontal unlike that of a stagnant liquid. However, there exists an
 upper limit to the slope of the top
surface, and the angle between this maximum slope and the
horizontal is known, for non-cohesive material, as the Coulomb critical angle $\tm$. Above this angle,
 the material becomes
unstable, and an avalanche at the surface might occur. The Coulomb angle is related to the friction
properties through $\tan \tm = \mu_{i}$ where $\mu_{i}$ is an internal friction
coefficient\C{Nedderman}.

As of today, the physical picture associated with the onset of the
avalanche is still obscure. One could imagine a local scenario in
which the dislodgement of some unstable grains leads by
amplification to a global avalanche (see for instance\C{RajchenbachCargese}). Alternately, one can think
of a delocalized mechanism\C{PGGJapon}, in which a thin slice of material is
destabilized and starts to slide as a whole. In the present paper,
we will focus on the latter point of view.

It has been recently suggested\C{PGGJapon} that the thickness of
the initial gliding layer should be of the order of $\xi$, the mesh size of the contact
force network\C{Radjai1,Radjai2,Liu}. For simple grain shapes (spheroidal), one expects
$\xi \sim$ 5--10 grain diameter $d$. The angle at which the avalanche
process actually starts is of the order of $\tm + \xi/L$, where $L$ is the size of the free surface.
At the moment of onset, our picture is that this initial layer starts to slip,
and is rapidly fluidized by the collisions with the underlying heap, therefore generating a
layer of rolling grains on the whole surface.

 Now that we have proposed a description of the initial situation, we may turn to the model scheme
 accounting for the further evolution of the avalanche.

\subsection{Saturation effects for thick avalanches}

Some years ago, Bouchaud, Cates, Ravi Prakash and Edwards
introduced a model to describe surface flows of granular
materials\C{Bouchaud94}. The model assumes a rather sharp
distinction between {\it immobile} particles and {\it rolling}
particles and, accordingly, introduces the following two important
physical quantities (see Fig.~\ref{BcrePicture}): the local height
of immobile particles $h(x,t)$ (where $x$ denotes the horizontal
coordinate\C{note1} and $t$ the time), and the local amount of
rolling particles $R(x,t)$.

The time evolution of $h(x,t)$ is written in the form
\begin{equation}
\label{Bcreh}
\frac{\partial h}{\partial t} =  \gamma R (\theta_n - \theta)
\end{equation}
where $\theta \simeq \tan(\theta) = {\partial h}/{\partial x}$ is
the local slope, $\gamma$ a characteristic frequency and $\theta_n$
the neutral angle of grains at which erosion of the immobile grains
balances accretion of the rolling grains. For the rolling
particles, Bouchaud and co-workers wrote a convection-diffusion
equation\C{Bouchaud94} that was later simplified by de Gennes
as\C{PGGCras95}
\begin{equation}
\label{BcreR}
\frac{\partial R}{\partial t} = v \frac{\partial R}{\partial x}
                                - \frac{\partial h}{\partial t}
\end{equation}
where $v$ is the downhill typical velocity of the flow, and is
assumed to be constant.

According to the Bouchaud-Cates-Ravi Prakash-Edwards (BCRE) model,
${\partial h}/{\partial t}$ is linear in $R$ [see Eq.\R{Bcreh}].
This is natural at small $R$, when all the rolling grains interact
with the immobile particles. But as explained in
Refs.\C{PGGCoursCdF97,Boutreux98}, this cannot hold when $R$
becomes larger than a given {\it saturation length} $\xi'$, since
the grains in the upper part of the rolling phase are no longer in
contact with the immobile grains. The length  $\xi'$ is expected to
be of the order of a few grain diameters $d$\C{note2}. This led
Boutreux, Rapha\"el, and de Gennes to propose\C{Boutreux98} a
modified {\it saturated} version of the BCRE Eq.\R{Bcreh}, valid
for thick surface flows and of the form
\begin{equation}
\label{BcrehSature}
\frac{\partial h}{\partial t} =  v_{uh} (\theta_n - \theta) \hspace{10mm} (R \gg \xi')
\end{equation}
where $v_{uh}$ is defined by $v_{uh} \equiv \gamma \, \xi'$. The constant $v_{uh}$ has
 the dimensions of a velocity.

The description of thick avalanches modelized by Eq.\R{BcrehSature}
was discussed in Ref.\C{Boutreux98}.
 However, one might encounter situations where the local amount $R$ of rolling particles is rather
 large except in some regions of space where it takes values smaller than $\xi'$. For such cases,
various `generalized' forms of the BCRE equations valid both in the
large and small $R$ limit, and able to handle intermediate values
have been proposed\C{PGGCoursCdF97,StopFlow,BouchaudCargese}.
 As we will be concerned only
with thick flows, we will henceforth use the saturated
form\R{BcrehSature}.

\subsection{Velocity profiles in thick flows}
\label{profile}

We now consider the hypothesis made in Eq.\R{BcreR} that the
downhill typical convection
 velocity of the rolling grains $v$ is constant. As a matter of fact, $v$ might vary for two
 reasons.

First, $v$  depends on the local slope $\partial h/ \partial x$ of
the static bed, reflecting that the mean convection velocity should
 increase as the sandpile is further tilted. However, in the situation we are going to consider,
 the slope should never depart from $\tn$ by more than a few degrees, so that the variations of $v$
originating in this may reasonably be taken to be negligible.

Second, $v$ might as well depend on the local amount of rolling
particles $R$. This dependence is quite natural, since as soon as
the thickness of the flow exceeds a few grain diameters,
 one would expect
 a velocity gradient perpendicular to the flow to establish. Such a possibility
was already considered by Bouchaud {\it et al.}\C{BouchaudCargese},
but, to our
 best knowledge,
 not further studied.  We think that taking this velocity gradient into account does lead to an
 improvement of the model description of avalanches. In the forthcoming sections we will analyse
the physical consequences of this modification.

If analyticity is assumed, we can expand $v(R)$ in powers of $R$,
and considering only the first
 two contributions to be significant, we write:

\begin{equation}
\label{DLv}
v(R)=v_{0} + \g R.
\end{equation}
with $\g$ a constant, homogeneous to a shear rate, and $v_{0}$ a
constant velocity.

When $R$ becomes small, Eq.\R{DLv} tells us that $v(R)$ becomes
constant [$v(R) \rightarrow v_{0}$]. Physically this velocity
should correspond to the typical convection velocity of a single
grain on a bed of immobile grains. For simple grain shapes
(spheroidal) and average levels of inelastic collisions, one
expects this velocity $v_{0}$ to scale as $(gd)^{1/2}$ (where $g$
is the gravity)\C{PGGCras95}. Similarly,
 the shear rate $\g$ is expected to scale as $(gd)^{1/2}/d \sim (g/d)^{1/2}$\C{note6}. We can therefore
 rewrite Eq.\R{DLv}:

\begin{equation}
\label{vdefinitif}
v(R)=\g (R + d).
\end{equation}
We note that $v_{0}$ becomes negligible compared to $\g R$ as soon
as $R$ exceeds a few grain diameters.

 In our approach, the typical velocity $v(R)$ depends \mbox{{\it linearly}} on the
 local rolling height $R$ [Eq.\R{vdefinitif}]. Such a form is in part motivated by the recent work
of Douady {\it et al.}\C{Douady} (see also
Section~\ref{Conclusion}). It is also supported by the
 experimental results of Rajchenbach {\it et al.} who carried measurements in a rotating
drum\C{DuranMRS,RajchenbachCargese}. These authors  have found
linear velocity profiles in the
 surface flow, with a shear rate $\g$ independent of the thickness
 of the flow. However, in other experiments of chute flows carried on rough inclined
planes, Anzaza {\it et al.}\C{Azanza} and Pouliquen\C{Pouliquen}
observe that the mean velocity (averaged on cross-sections)
 scales as a power-law of the thickness with an exponent about 3/2. In the following we will
 mainly focus on the linear form\R{DLv}, since it allows us to give explicit analytical solutions,
 and shall discuss the changes that are to be brought in the case of a power-law velocity
in Section~\ref{sectionRmax}.

In the next section, we will derive the governing equations from
the saturated BCRE equations
 and the above considerations on the velocity profile inside the flow.

\subsection{Governing equations}

We may define a reduced profile $\hti$, deduced from $h$ by
substracting the `neutral' profile
 $\tn \, x$:

\begin{equation}
\label{Defhtilde}
\hti(x,t) \equiv h(x,t) - \tn \, x.
\end{equation}

Using Eqs\R{BcreR},\R{BcrehSature},\R{vdefinitif} and\R{Defhtilde},
we easily obtain the following
 system:

\begin{eqnarray}
\label{eqh}
\frac{\partial \hti}{\partial t} & =& - \, \vuh \frac{\partial \hti}{\partial x} \\
\label{eqR}
\frac{\partial R}{\partial t} & = & \g (R+d) \frac{\partial R}{\partial x}
   + \vuh \frac{\partial \hti}{\partial x}.
\end{eqnarray}

In our approach, Eqs.\R{eqh} and \R{eqR} are the governing
equations for surface avalanches displaying linear velocity
profiles.

An important point is that we must have $R > 0$ for Eqs.\R{eqh}
and\R{eqR} to be valid. If we reach
 $R=0$ in a certain spatial domain, then Eq.\R{eqh} must be replaced in that domain by
$\partial \hti / \partial t = 0$.

\section{Application to the simple case of an open system}

\subsection{Physical situation}

We will now solve Eqs.\R{eqh} and \R{eqR} in the following simple
situation: we consider a cell, of dimension $L$, partially filled
with monodisperse grains of diameter $d$, as shown on
 Fig.~\ref{figOpenCell}.
 The heap has an initial uniform slope $\tm$, the Coulomb angle of the material. The origin of the
 $x$ axis is taken at the
 bottom of the cell, and the orientation of the axis is such that the slope of the heap is positive.

 We now consider that an avalanche has just started in the cell (see Section~\ref{Onset}), so
that we have at time $t=0$ a layer of rolling grains in the whole
cell, of thickness $\sim \xi$ greater than the saturation length
$\xi'$. We may thus use the saturated equations\R{eqh} and \R{eqR}
from the beginning of the avalanche.

As the rolling population will rapidly grow and become independent
of the initial thickness
 $\xi$ (for $\xi$ small), we can as well consider the initial condition on $R$ to be:

\begin{equation}
\label{condinitR}
R(x,t=0)=0.
\end{equation}

We also know the initial value of $\hti$:

\begin{equation}
\label{condinith}
\hti (x,t=0)= (\tm - \tn) \, x \equiv \eta \, x,
\end{equation}
where $\eta$ is defined as the (positive) difference between the
Coulomb angle and the neutral angle.

We have additional conditions in our system, due to the boundaries.
At the top of the cell,
 there is no feeding in rolling species, so that we impose:

\begin{equation}
\label{condtop}
R(x=L,t)=0 \hspace{10mm} \mbox{at any time } t \geq 0.
\end{equation}
Another condition arises from the fact that grains fall off the
cell at the bottom and cannot accumulate there:

\begin{equation}
\label{condbottom}
\hti (x=0,t)=0 \hspace{10mm} \mbox{at any time } t \geq 0.
\end{equation}

\subsection{Uphill wave in the static phase}

Equation\R{eqh} can be readily solved along with
conditions\R{condinith} and\R{condbottom} to give:

\begin{equation}
\label{h}
\hti (x,t) = \eta \vuh H(x - \vuh t) \,x \mbox{ \ \ for } 0 \leq\ x \leq L,
\end{equation}
 where
$H$ denotes the Heavyside unit step function [$H(u)=1$ if $u>0$,
$H(u)=0$ otherwise]. This result
 corresponds to the uphill propagation (at constant speed $\vuh$) of a surface wave on the
static phase. Let us call $x_{uh}(t)$ the time-dependent position
of the wavefront, given by:

\begin{equation}
\label{Defxuh}
x_{uh}(t)=\vuh t.
\end{equation}
(where the subscript $uh$ stands for `uphill').

The wave starts from the bottom of the cell at time $t=0$ and
reaches the upper end at time $t_{2}$
 defined by:

\begin{equation}
\label{t2}
t_{2} \equiv L/\vuh.
\end{equation}

At a given time $t$ (smaller than $t_{2}$), the profile of the
static phase can be described as
 follows: ahead of the wavefront
\mbox{[$x_{uh}(t) \leq x \leq L$]}, the profile is linear and the slope is the initial angle $\tm$
(since $\hti=\eta \vuh$). Behind the wavefront [$0 \leq x \leq
x_{uh}(t)$], the slope has decreased and reached the neutral angle
$\tn$ ($\hti=0$) (see Fig.~\ref{figh}). For times $t \geq t_{2}$
the
 slope of the static phase inside the cell is uniformly equal to the final value $\tn$, which is
 thus the angle of repose of our specific open cell system~\C{note3}.

\subsection{Downhill convection of rolling grains}

Substituting Eq.\R{h} into the evolution equation\R{eqR} for $R$
gives:

\begin{equation}
\label{equadiffR}
\frac{\partial R}{\partial t}  - \g (R+d) \frac{\partial R}{\partial x}
  = \eta \vuh H(x - \vuh t)
\end{equation}

Eq.\R{equadiffR} is a non-linear convection equation. The rolling
species are thus convected downhill, with a  convection velocity
dependent on the local rolling thickness $R$. In the spatial region
$x > \vuh t$, the right-hand side (which couples the evolution of
$R$ to that of $\hti$) plays the role of a source term, leading to
an amplification of the avalanche.
 On the contrary, for $x \leq \vuh t$, the right-hand side goes to zero, so that the material
 flowing through the surface $x=\vuh t$ from uphill is simply convected, without amplification
 nor damping.

Equation\R{equadiffR} can be solved analytically by using the {\it
method of characteristics}\C{characteristics}, which utilizes the
property that
 certain types of partial differential equations reduce to a set of
 ordinary differential equations along particular lines, known as the characteristic curves.
 For more details on this method and on its application in the case of Eq.\R{equadiffR}, see
 Appendix.

\subsection{Propagation of boundary effects in the cell}

Before we go to the precise solutions, we can try to get some
physical insight of the way the avalanche is going to develop.
 The global shape of the rolling phase at different moments during the avalanche is of course very
 dependent on the boundary condition\R{condtop} for $R$ in the cell, but also
on the condition\R{condbottom} for $\hti$, since the evolution of
$\hti$ and $R$ are coupled.

However, the effects of these boundary conditions cannot spread
over the entire cell instantly after
 the beginning of the avalanche, and shall propagate with finite
 velocities. We then expect the progression of
these boundary effects (one could say, the propagation of the
`information' on the boundaries) to control the evolution of
 both the rolling and the static phase. For instance, in the case of the static profile $\hti$,
Eq.\R{h} tells us that the bottom boundary condition\R{condbottom} [$\hti(x=0,t)=0$
at any time $t$] brings progressively $\hti$ to zero everywhere in the cell, and also, that
the propagation proceeds with a velocity $\vuh$.

Hence, we expect that the description of the avalanche should
 naturally split up in different temporal `stages', according to the degree of extension of the
different
 boundary effects, and that the cell should divide in several `regions', according to whether
it is under
 the influence of the top boundary condition or the bottom, or both, etc. This shall become clear
 as we will now go into the precise description of the avalanche.

\section{Unfolding of the avalanche}

\subsection{Stage I: The avalanche grows to maturity}
\label{stageI}

This stage starts at $t=0$ with the beginning of the avalanche.
From the above considerations, we know that the boundary effects start to propagate
with finite velocities from both ends of the cell. We can therefore define `propagation fronts'
for these effects: we call $x_{dh}(t)$ the position of the front originating in
the boundary condition at the {\it top} of the cell [the subscript $dh$
means that the motion of this front is downhill], and $x_{uh}(t)$ the corresponding `uphill'
front, originating in the {\it bottom} boundary condition, and that we already defined
earlier as $x_{uh}(t)=\vuh t$ [Eq.\R{Defxuh}].
Figure~\ref{figStageI}-a presents a typical picture of the situation during Stage
I, where the fronts, after leaving their respective cell ends, move in opposite directions and
 one toward the other.
As a consequence, they shall finally meet at a certain time, that we hereafter denote $t_{1}$.
This time $t_{1}$ defines the end of what we call `Stage I' (which is thus characterized as the time
 interval $0 \leq t \leq t_{1}$), and the beginning of `Stage II' (described in next section).

The relative positions of the fronts naturally define three spatial
regions in the cell (Fig.~\ref{figStageI}-a). To the left of $x_{uh}$, the effects
of the bottom boundary condition [Eq.\R{condbottom}] are
predominant. We call this region the {\it bottom} region. We
remark that it is constantly extending uphill during Stage I
[following the motion of $x_{uh}(t)$]. To the right of
$x_{dh}(t)$, we define the {\it top} region, which extends downhill, and where the
evolution of the avalanche is controlled by the upper end condition [Eq.\R{condtop}].
Finally, between those two regions remains a {\it central} region, where none of the boundary
effects can yet be felt. This last region shrinks during Stage
I, and ultimately disappears at time $t=t_{1}$ when the bottom and
top region connect. We now describe the precise evolution of the
avalanche, region by region [we will only give the form of the rolling
amount $R(x,t)$, since $h(x,t)$ is already known from Eq.\R{h}].

\subsubsection{Top region}

From the above definition, the top region corresponds to the spatial domain
$x_{dh}(t) \leq x \leq L$. Within this domain, Eq.\R{equadiffR} reads:

\begin{equation}
\label{EquaDiffRtop}
\frac{\partial R}{\partial t}  - \g (R+d) \frac{\partial R}{\partial x}
 = \eta \vuh
\end{equation}
[since $x > x_{dh}(t) > x_{uh}=\vuh t$].

Solving this equation with the boundary condition \mbox{$R(x=L,t)=0$}
(see Appendix for details) gives the expression of $R$ valid in this
region:

\begin{equation}
\label{R1top}
R(x,t)= -d + \sqrt{d^{2} + 2(L-x) \frac{\eta \vuh}{\g}}.
\end{equation}
We also obtain the precise position of the `downhill front'
$x_{dh}(t)$:

\begin{equation}
\label{Defxdh}
x_{dh}(t) = L + \frac{\g}{2 \eta \vuh} d^{2} -
  \frac{1}{2} \g \eta \vuh \left(t + \frac{d}{\eta \vuh} \right)^{2}.
\end{equation}

Thus, according to Eq.\R{R1top}, $R$ has a stationary shape (independent of
time), but on a domain that extends downhill with time. Interestingly, we note that
the motion of $x_{dh}(t)$ is uniformly accelerated throughout the stage.
This is a direct consequence of the non-linearity in Eq.\R{equadiffR}.

\subsubsection{Central region}

In the central region, the boundary conditions have no influence on
the evolution of $R$ and $\hti$. The central region is hence spatially defined by
$x_{uh}(t) \leq x \leq x_{dh}(t)$, and shrinks at both ends to disappear at the end
 of the stage.

The evolution equation for $R$ is the same as in the top region [Eq.\R{EquaDiffRtop}], but now we
must impose the initial condition\R{condinitR} (no boundary condition). The solution (see Appendix) writes:

\begin{equation}
\label{R1central}
R(x,t)  = \eta \vuh t.
\end{equation}

In the central region the rolling phase grows linearly with time and
uniformly in space, thus forming a plateau (see
Fig.~\ref{figStageI}-b). This constant growth rate is a consequence of
the saturated form
of the BCRE equations we have used [Eq.\R{BcrehSature}]. The uniformity of the solution, on the
other hand, stems from the fact that since none of the boundaries is at work, and
since the initial static profile was uniform, the central region behaves like an infinite medium
for which translational invariance is to be obeyed.

We finally remark that solutions\R{R1top} and\R{R1central} connect
continuously
 at $x=x_{dh}(t)$.

\subsubsection{Bottom region}

The bottom region is controlled by the bottom boundary condition,
and spreads over the spatial
 interval \mbox{$0 \leq x \leq x_{uh}(t)$}. In this region the evolution equation for $R$ displays
 no more amplification (because $x < x_{uh}(t)=\vuh t$):
\mbox{$\partial R/\partial t  - \g (R+d) \: \partial R/\partial x = 0$}.

Since there is no constraint on $R$ at the bottom of the cell, the
condition on $R$ is given
 by the physical assumption that it should be continuous across the border of the central and bottom
 regions, i.e. $R(x=x_{uh}(t),t)= \eta \vuh t$ for $t \geq 0$.

This leads to the following expression for $R$:

\begin{eqnarray}
\label{R1bottom}
\lefteqn{R(x,t) = -\; \frac{1}{2} \left(d + \frac{\vuh}{\g} - \eta \vuh t \right) }\nonumber
 \hspace{7mm} \\
 & & + \; \frac{\vuh}{2 \g} \sqrt{ \left(\frac{\g d}{\vuh} +1 - \g \eta\ t \right)^{2}
 + 4 \frac{\g \eta}{\vuh} (x + \g d t) }.
\end{eqnarray}

In this region also the height of rolling grains increases with time, due to
an increasing input of material at the frontier with the central region.

\subsubsection{Derivation of time $t_{1}$}

Stage I ends when the top and bottom regions meet, at $t=t_{1}$
defined by $ x_{uh}(t=t_{1})=x_{dh}(t=t_{1})$. Using
Eqs.\R{Defxuh} and\R{Defxdh}, we easily obtain

\begin{equation}
\label{t1}
\hspace{-0.25 cm} t_{1}= -\left(\frac{d}{\eta \vuh} + \frac{1}{\g \eta} \right)
                                                + \sqrt{ \left(\frac{d}{\eta \vuh} + \frac{1}{\g \eta} \right)^2
                                                      + \frac{2 L}{ \g \eta \vuh}}.
\end{equation}

As will be shown later, the maximum thickness of the avalanche,
$R_{max}$, is actually reached for $t=t_{1}$ and
$x=x_{uh}(t=t_{1})=x_{dh}(t=t_{1})$. We can clearly see on
Fig.~\ref{figStageI}
 that the $R$-profile at time $t=t_{1}$
 displays a cusp. As the prediction of the maximum amplitude $R_{max}$ is an important result of our analysis, we
shall devote Section~\ref{sectionRmax} to it, and defer the
analytical derivation of $R_{max}$
 and its application to physical examples until there.

Fig.~\ref{figStageI}-b presents successive `snapshots' of the rolling
phase profile during Stage I.

\subsection{Stage II: The static profile reaches its final state}

Stage II starts at $t=t_{1}$. At time $t_{1}$, the two `propagation fronts' of the boundary effects
$x_{uh}(t)$ and $x_{dh}(t)$ pass each other, and then pursue their respective motions
towards the opposite cell edge. Figure~\ref{figStageII}-a illustrates this situation.
As in Stage I, it appears that the cell is naturally divided in three spatial regions: a top region
[defined spatially as $x_{uh}(t) < x \leq L$], under the sole influence
 of the upper edge of the cell; a bottom region \mbox{[$0 \leq x < x_{dh}(t)$]}, under the
influence of the bottom edge; and finally, a central region [$x_{dh}(t)
\leq x \leq x_{uh}(t)$], where in contrast with Stage I, the effects of {\it both} boundaries now combine.
As another difference with the situation described in Stage I, the top and bottom regions progressively shrink,
whereas the central one grows in extension (Fig.~\ref{figStageII}-a).

Due to their motion, the fronts $x_{dh}$ and $x_{uh}$ are bound to reach, sooner or later, the bottom
and top end of the cell (respectively). At time $t_{2}$ [Eq.\R{t2}], the uphill front reaches the upper
limit of the cell ($x_{uh}=L$). The static profile is then in its relaxed final state, with a uniform slope $\tn$.
This is the end of Stage II, which is thus defined as the time interval $t_{1} < t \leq t_{2}$.
In most cases, as is discussed below, we expect the downhill front
to reach the bottom edge {\it before} $t=t_{2}$.

\subsubsection{Top region}

In this region, we have $x>x_{uh}$, so that the evolution equation\R{EquaDiffRtop} still holds, and we still
have to solve with respect to the upper boundary condition of Eq.\R{condtop}. Therefore, as in Stage I, $R$
is given  by Eq.\R{R1top} [but now, the lower limit of the
domain on which this solution is valid is $x_{uh}(t)$, not $x_{dh}(t)$]. This top
region shrinks, until finally disappearing when the uphill
front reaches the upper end of the cell ($t=t_{2}$).

\subsubsection{Central region}

In this part, since $x \leq x_{uh}(t)$, there is no amplification
of the rolling amount, so that the right-hand side of the evolution
equation of $R$ vanishes:
 \mbox{$\partial R/\partial t  - \g (R+d) \: \partial R/\partial x = 0$}.
Now, we further impose that $R$ shall be continuous at the border with the
top region, i.e.:

\[
R(x=x_{uh}(t), t) =  -d + \sqrt{d^{2} + 2(L-x) \frac{\eta
\vuh}{\g}}.
\]

Solving these two equations together leads to look for $R$ as one of the roots of
the third-degree equation \mbox{$R^3 + a_{2}R^2 + a_{1}(t)R + a_{0}(x,t) = 0$},
with the following coefficients:

\begin{eqnarray*}
a_{2} & = & \left(\frac{\vuh}{\g} + 3d \right)  \\
a_{1}(t) & = & 2 \frac{\vuh}{\g} \left(d + \frac{\g d^2}{\vuh} -
\eta (L-\vuh t) \right) \\ a_{0}(x,t) & = & - 2 \eta
\frac{\vuh}{\g} \left( \frac{\vuh}{\g} (L- x) + 2d(L- \vuh t)
\right).
\end{eqnarray*}

The solution, given by Cardano formulas\C{Bronshtein}, writes:

\begin{equation}
\label{R2central}
R(x,t)= -\frac{a_{2}}{3} + S - \frac{Q}{S},
\end{equation}
with the auxiliary quantities\C{note4}:
\begin{eqnarray*}
S &\equiv& \sqrt[3]{P + \sqrt{D}} \\ D &\equiv& Q^3 + P^2 \\ Q &\equiv&
\frac{1}{9}(3a_{1}-a_{2}^2) \\ P &\equiv& \frac{9 a_{2} a_{1} - 27 a_{0}
-2a_{2}^3}{54}.
\end{eqnarray*}

We saw in the previous section that the crest of the avalanche
$R_{max}$ appeared at the end of Stage I.
 What happens to this crest during Stage II? It is easy to prove
that the crest remains located on the downhill front $x_{dh}$.
Besides, $x_{dh}$ now moves at constant speed (in contrast with
Stage I where it accelerated):
\mbox{$x_{dh}(t)=\vuh t_{1} - \g (R_{max}+d)(t-t_{1})$}.

We can also prove that the height of the crest remains constant (equal to $R_{max}$) as
it travels downhill, until it finally comes out of the cell.
The exit time $t_{exit}$ of this crest $R_{max}$ is obtained by
 solving $x_{dh}(t)=0$:

\begin{equation}
\label{texit}
\hspace{-0.2 cm} t_{exit}=t_{1}+ \frac{\vuh t_{1}}{\g (R_{max}+d)}=t_{1} \left(1 +
\frac{\vuh}{\g (R_{max}+d)} \right).
\end{equation}

\subsubsection{Bottom region}

The bottom region is defined as the region where \mbox{$0 \leq x
\leq x_{dh}(t)$}. The evolution equation for $R$ is the same as in the central region, and we impose continuity
at the border with the central region. We find that $R(x,t)$ is given by Eq.\R{R1bottom} as in Stage I.
Physically, in this region, we simply observe the convection of what was left in the
bottom region at the end of Stage I.

The bottom region disappears when the downhill front reaches
the bottom end: $x_{dh}(t)=0$, that is by definition at time $t=t_{exit}$ [Eq.\R{texit}]. To
determine precisely the subsequent evolution of the avalanche, we
must discuss whether the disappearance of the bottom region occurs before the end of Stage II or not
(that is, whether $t_{exit} \leq t_{2}$ or $t_{exit} \geq t_{2}$).
Using Eq.\R{texit}, we form the
 ratio:
\[
\frac{t_{exit}}{t_{2}}= \frac{t_{1}}{t_{2}} \left(1 + \frac{\vuh}{\g (R_{max}+d)} \right).
 \]

Provided that the cell dimension $L \gtrsim 100d $ and
that $\vuh \sim \g d$ (these requirements being usually satisfied
for common experiments), we have $\g R_{max} \gg \vuh$,
 and $t_{1}/t_{2} \ll 1$. Hence we generally expect $t_{exit} \ll t_{2}$, and, consequently,
  as claimed earlier, in most cases the bottom region disappears before the end of Stage II.

To resume, during Stage II, the central region extends both
downhill and
uphill with constant (though different) velocities at each end, progressively invading the  whole
cell. It reaches the bottom edge at time $t_{exit}$ (in situations where $t_{exit} \ll t_{2}$),
then the upper edge at time $t_{2}$.  At the end of Stage II, the central region occupies
the entire cell, and $R(x,t=t_{2})$ is everywhere given by Eq.\R{R2central}. We present successive
calculated `snapshots' of the rolling amount $R$ during this stage in Fig.~\ref{figStageII}-b.

\subsection{Stage III: The last grains are evacuated}

This stage lasts from $t=t_{2}$ until the end of the avalanche, at
$t=t_{end}$. Both fronts have reached the edges of the cell
at the end of Stage II ($x_{dh}=0, x_{uh}=L$), and there is only one region (see Fig.~\ref{figStageIII}-a).
Moreover, as $t>t_{2}$, the slope of the static part is everywhere $\tn$ and
no amplification of the rolling grains can take place; the rolling phase
is simply convected downwards.
We now have to solve the evolution equation \mbox{$\partial R/\partial t  - \g (R+d) \: \partial R/\partial x =
0$}, with respect to the initial condition that Stage III evolves from what has been left by
Stage II [i.e. $R(x,t=t_{2})$ as given by Eq.\R{R2central}].

Solving with the method of characteristics gives the following
 implicit solution

\begin{equation}
\label{R3}
R(x,t) = R(\xi,t_{2}),
\end{equation}
where
\begin{equation}
\label{xi}
\xi= x \, - \, \g \left[\, R(\xi,t_{2}) +d \, \right] (t-t_{2})
\end{equation}
(and $0 \leq \xi \leq L$).

The physical interpretation of these equations is actually very
simple: Eq.\R{R3} states that
 the quantity of rolling species found in $x$ at time $t$ was previously located in $\xi$ at the
 beginning of Stage II $(t=t_{2})$. Equation\R{xi} gives a determination of this initial position
 $\xi$, by stating that from $t_{2}$ until the considered instant $t$, the quantity of grains
 moved with a constant speed $\g[\, R(\xi,t_{2}) +d \,]$, dependent on the local height. In
 other words, during Stage III, the $R$-profile left by Stage II is convected downhill, but
each vertical slice rolls with its own velocity, which is a function
of its height. The
 grains that were near the top edge of the cell at the end of Stage II are convected the
 most slowly, since there $R$ was close to zero. The profile inherited from Stage II thus dilates
 upon rolling, under the effect of velocity inhomogeneities (Fig.~\ref{figStageIII}-b)\C{note5}.

The last grains to fall off the cell are those that leave the top
end of the cell at the beginning of Stage III, at time $t_{2}$.
Since at the top edge we have $R=0$, these grains move
 with a constant speed $v_{0}=\g d$. At time $t$, they are located at $x_{last}(t) = L - \g d \: (t - t_{2})$,
 and the avalanche is extinct uphill: $R=0$ for $x > x_{last}(t)$.

Finally, the avalanche ends when the last grains reach the bottom
limit of the cell ($x_{last}=0$),
 that is at time $t_{end}=t_{2} + L/(\g d)$.

\section{Discussion and simple checks}

\subsection{Predictions for the maximum amplitude of the avalanche}

\label{sectionRmax}

\subsubsection{Linear velocity profile}

Up to now, we have focused on flows displaying linear velocity
profiles. For such flows, as we saw in Section~\ref{stageI}, the
avalanche reaches its maximum amplitude $R_{max}$ at the end
 of Stage I, at time $t=t_{1}$ [Eq.\R{t1}]. The exact analytical expression of $R_{max}$ is easily
  found by using Eq.\R{R1central} at time $t_{1}$ (that is, the value of $R$ given by the central region
   at the very moment it disappears):

\begin{equation}
\label{Rmax}
R_{max} = -d - \frac{\vuh}{\g}
        + \sqrt{ \left(d + \frac{\vuh}{\g} \right)^2 + \frac{2 L \eta \vuh}{\g}}.
\end{equation}

For large values  of L, $R_{max}$ scales as:

\begin{equation}
\label{RmaxScaling}
R_{max} \sim \sqrt{2 \eta \frac{\vuh}{\g} L},
\end{equation}
that is, as the square root of the system size $L$.

Let us give a couple of numerical applications of this last
expression. For the case of a standard laboratory experiment, with
$L=1$ m, $d=1$ mm, $\vuh/\g = 3d$ and $\eta \sim 0.1$ rad, we find
\mbox{$R_{max}=2.45$ cm}. In the case of a system at the scale of a
 desert dune,
 made of fine sand, we take $L= 10$ m, $d=0.2$ mm and, with others parameters unchanged, we get
 $R_{max}=3.46$ cm. One has to notice that $R_{max}$ is quite small, even for large systems as
 a sand dune.

It is interesting to contrast this result with the work of Boutreux
{\it et al.}\C{Boutreux98}, who carried the same calculation in an
open cell configuration, but with a constant downhill convection
velocity $v(R) \sim v_{0}$ [instead of Eq.\R{DLv}]. They found
$R_{max} \sim \eta L$.
 For the two above examples, this formula leads to maximum amplitudes of respectively $10$ cm and
 $1$ m. The effect of the velocity gradient is thus to considerably limit the amplitude of
avalanches, especially for large systems.

\subsubsection{Generalization for a power-law dependency}

In the beginning of this article, we quoted the work of Azanza {\it
et al.}\C{Azanza} and of Pouliquen\C{Pouliquen} who find
 that the average speed of a chute flow of granular material on a rough plane is related to its
 thickness through a power-law relation $v(R) \sim \g R^{\alpha}$ with $\alpha$ close to 3/2.
 However, as pointed out by Pouliquen\C{Pouliquen}, the influence of the rough underlying bed
 plane on the rheology of chute flows is complex and not clearly understood, and might not be comparable to
 situations where the flow occurs on a {\it free} granular bed as has been considered in this paper.
 Since the question is still open, we will here  present an intuitive derivation of
 $R_{max}$ valid  for any power-law \mbox{(undetermined exponent $\alpha$)}. To check the
 validity of this simple derivation, we first present it in the linear case $\alpha = 1$, the
 generalization being then straightforward.

Let us consider a point initially at the top edge of the cell. At $t=0$, it starts being swept along
 by the granular flow and we assume that this point travels with the local surface velocity of the flow $v= \g R$.
  We are now interested in the temporal evolution of the rolling height $R$ {\it at this travelling point}, which shall
   be computed from the Lagrangian derivative $dR/dt = \partial R/\partial t + v \: \partial R/\partial x$.
As long as the amplification process takes place, we have with the use of Eq.\R{equadiffR}: $dR/dt = \eta \: \vuh$.
This implies

\begin{equation}
\label{Ramp}
R(t) = \eta \vuh t.
\end{equation}

Hence, $R(t)$ at the travelling point increases with time, as long as the amplification process
lasts. After the amplification has stopped, $R$ at the travelling point keeps
constant (since $dR/dt = 0$). Thus, $R$ reaches its maximum value $R_{max}$
at the end of the amplification.
 Let us call $t_{amp}$ the duration of the amplification.
We compute $t_{amp}$ in the following way: the distance that the
travelling
 point goes over during the amplification is of order $\sim L$, so that $t_{amp}$ must verify:

\begin{eqnarray*}
L &\sim & \int_{0}^{t_{amp}} dx = \int_{0}^{t_{amp}} v \, dt  \\
\mbox{i.e.} \hspace{4mm} L & \sim & \int_{0}^{t_{amp}} \g \eta\ \vuh t \, dt =
                                                         \frac{1}{2} \g \eta \vuh {t_{amp}}^{2},
\end{eqnarray*}
[in this calculation, we used $v \sim \g R(t)$]. We finally find:
 \mbox{$t_{amp} \sim \sqrt{2 L/(\g \eta \vuh)}$}.

Inserting this last expression into Eq.\R{Ramp} gives the value of $R_{max}$:
\begin{equation}
R_{max} \sim \sqrt{2 \eta \frac{\vuh}{\g} L}.
\end{equation}

This is exactly Eq.\R{RmaxScaling} found analytically, which was
itself the limit of the complete expression of $R_{max}$ [Eq.\R{Rmax}] for
large values of L (greater than a hundred $d$), and $\vuh$ of order
$\g d$.

The strongest assumption in the above simple derivation is that the
 amplification takes place over a \mbox{distance $\sim L$}. Rigorously, this distance is
\mbox{$L - x_{dh}(t=t_{1})$}; but what makes our simple derivation successful is that the position of
 the downhill front at time $t_{1}$, $x_{dh}(t=t_{1})$, is generally quite close to zero (for $L$
 greater than a hundred $d$ and $\vuh$ of order $\g d$).

We may now generalize the above results to a power-law dependency
of the velocity $v(R) \sim \g R^\alpha$. The same derivation leads
us to the result:

\begin{equation}
\label{Rgeneral}
R_{max} \sim \left( (\alpha +1) \frac{\eta \vuh}{\g} \; L
\right)^{1/(\alpha + 1)}
\end{equation}
Note that $R_{max}$ diminishes as $\alpha$ increases.

In particular, for $\alpha=3/2$, Eq.\R{Rgeneral} can be rewritten
as $R_{max} \sim ( 5 \, \eta L \vuh / \, 2 \, \g )^{\: 2/5}$.

\subsection{Possible experimental checks}

The loss of material at the bottom edge of the cell might be measured experimentally, and could be
compared to the following theoretical prediction. This loss corresponds to the flow rate at the bottom
of the cell $Q(x=0,t)= \int_{0}^{R(x=0,t)} v(z) dz$, and is given by
\begin{equation}
\label{loss}
Q(x=0,t)=\frac{\g}{2} R(x=0,t)^{2} + \g d \: R(x=0,t),
\end{equation}
where $R(x=0,t)$ is given by Eq.\R{R1bottom} during Stages I and II, and by Eq.\R{R3} during Stage III.
Figure~\ref{figRoutput} shows the predicted shape of $Q(x=0,t)$ as a function of time (solid curve).
The curve displays a maximum at time $t=t_{exit}$, corresponding to the moment when the maximum amplitude
$R_{max}$ rolls out of the cell. The maximum flow rate is obtained by replacing $R(x=0,t)$ by $R_{max}$
in Eq.\R{loss}.

It is of interest to compare our prediction for the loss of
material with that of Boutreux {\it et al.}\C{Boutreux98}, who
assumed a constant downhill velocity $v$ in the rolling phase. This
comparison, however, requires some caution: in our approach, the
granular flow is characterized by a constant velocity gradient
$\g$, whereas, in Boutreux {\it et al.}, the description is based on
 a typical downhill convection velocity of the grains $v$ (see
Section~\ref{profile}). Figure~\ref{figRoutput} compares the
results of both approaches for the loss of material, assuming $v \simeq \vuh \simeq 3 \: \g d$
(see Ref.\C{Boutreux98}).

\subsection{Concluding remarks}
\label{Conclusion}

\subsubsection{Regions of small $R$}

We notice that, during the avalanche, we had several spatial zones
in the cell where R was close to 0 ($R<\xi'$), e.g. at the upper
edge of the cell, or at the end of the avalanche. Thus in these
zones, the use of the saturated Eqs.\R{eqh} and\R{eqR} is not fully
justified. In order to obtain a continuous description between the
saturated case and the thin one, we could use the interpolated
equations that have been proposed by de Gennes\C{PGGCoursCdF97}
and studied in a model case by Boutreux and  Rapha\"el in Ref.\C{StopFlow}:

\begin{eqnarray*}
\label{eqhInterpolee}
\frac{\partial \hti}{\partial t} & = & - \, \gamma \, \frac{R \xi'}{R+\xi'} \frac{\partial \hti}{\partial x} \\
\label{eqRInterpolee}
\frac{\partial R}{\partial t} & = & \g (R+d) \frac{\partial R}{\partial x}
 + \gamma \, \frac{R \xi'}{R+\xi'} \frac{\partial \hti}{\partial x}.
\end{eqnarray*}

The results of\C{StopFlow} show however that the physical behaviour
is not dramatically changed,
 and that the description in the zones of small R with saturated equations might be slightly wrong
 but qualitatively verified.

\subsubsection{Effects of polydispersity}

It is of common knowledge that real granular materials
 are generally intrinsically polydisperse. This may have drastic effects on the behavior of the flow,
 and capturing more precisely the physics of real avalanches would certainly suppose to take
polydispersity into account. However, the treatment of full
polydispersity is a difficult task. Yet,
 the BCRE equations have been extended to the case of binary
mixtures\C{BoutreuxdeGennes96,TheorieMakse}, and it could be
interesting to study the changes brought up in this case by a
velocity gradient in the flow.

\subsubsection{Domain of validity of the BCRE approach}

The general approach introduced by Bouchaud {\it et al.} to
describe surface flows is rather phenomenological, and as pointed
out by Bouchaud and Cates\C{BouchaudCargese}, we still lack
criteria to determine the range of physical situations to which it
can be successfully applied.
 In a recent work, Douady {\it et al.}\C{Douady} proposed a justification of
 the BCRE modelization on the basis of hydrodynamic
 conservation laws. According to these authors, the form of the BCRE equations should not remain
 invariant when different laws are chosen for the velocity profile in the flow. Douady {\it et al.}
 argue that only for a velocity linear in $R$ (or constant) shall the equations take the
 simple form of our equations [Eqs.\R{eqh} and\R{eqR}]; in other cases, they find that a
supplementary term coupling  $R$ and $h$ should add in Eq.\R{eqh}.
Certainly, more work needs to
 be done in this direction in order to exactly assert the domain of validity of the BCRE analysis.

\acknowledgements
We thank T. Boutreux, F. Chevoir, A. Daerr, \mbox{S. Douady}, J. Duran and
O. Pouliquen for oral and/or written exchanges.

\appendix
\section*{}

\subsection{Method for solving the evolution equation of $R$}

Eq.\R{equadiffR} is a first-order partial differential equation, of
the quasi-linear class, that is, linear in the first derivatives.
Such equations can be solved by the well-known method of
characteristics. See, for example, Ref.\C{characteristics}.

More specifically, we will solve Eq.\R{equadiffR} along
characteristic curves given in the parametric form $\{t(s), x(s),
R(s)\}$, with $s$ the parameter. The functions $t(s)$, $x(s)$ and
$R(s)$ are derived from the set of coupled ordinary differential
equations:

\begin{eqnarray}
\frac{dt}{ds} &=& 1 \nonumber \\
\label{eqcarac}
\frac{dx}{ds} &=& - \g (R+d) \\
\frac{dR}{ds} &=& \eta \vuh H(x - \vuh t) \nonumber.
\end{eqnarray}

By integration, one founds the equations for the characteristics
with unspecified integration constants. One then imposes the
boundary and/or initial conditions to identify these constants.

We here give the detailed calculations only for the first stage of
the avalanche. The derivations
 are separated into the different spatial regions that were defined earlier, and we will show how
 they naturally emerge from the derivations.

\subsection{Top region}

In this region, Eqs.\R{eqcarac} become:

\begin{eqnarray}
\frac{dt}{ds} &=& 1 \nonumber\\
\label{eqparam}
\frac{dx}{ds} &=& - \g (R+d) \\
\frac{dR}{ds} &=& \eta \vuh \nonumber.
\end{eqnarray}

We also use the boundary condition $R(x=L,t)=0$ for $t \geq
0$, which we parametrize with the parameter $\xi$. For
simplicity's sake, on each characteristic crossing the boundary
curve we arbitrarily choose the value of the parameter $s$ to be
zero at the crossing point. This determines the integration
constants to be:

\begin{eqnarray}
t(s=0) & = & \xi \nonumber \\
\label{eqinit}
x(s=0) & = & L \\ R(s=0) & = & 0 \nonumber.
\end{eqnarray}

Note that $\xi \geq 0$, since $t \geq 0$ (the experiment started at time $t=0$ on).

Solving for Eqs.\R{eqparam} together with Eqs.\R{eqinit} gives the
equations of the characteristic curves:

\begin{eqnarray}
\label{tparam1top}
t(s) & = & s + \xi  \\
\label{xparam1top}
x(s) & = & - \frac{1}{2} \g \eta \vuh \: s^{2} -\g d \: s + L \\
\label{Rparam1top}
R(s) & = & \eta \vuh \: s.
\end{eqnarray}

We now want to write the solution $R$ explicitly in terms of $x$
and $t$, so that we have to
 eliminate $\xi$ and $s$. Eq.\R{xparam1top} can be solved to give $s$ as a function of $x$, and
 replacing into Eq.\R{Rparam1top} brings the analytical solution:

\begin{equation}
\label{R1top_appendix}
R(x,t)= -d + \sqrt{d^{2} + 2(L-x) \frac{\eta \vuh}{\g}},
\end{equation}

which is the same as Eq.\R{R1top}.

We now have to verify the condition that $\xi \geq 0$. By combining Eq.\R{tparam1top} with \R{Rparam1top}
 this condition can be rewritten
as $t \geq R(x,t)/ \eta \vuh$. Replacing $R$ into this inequality [by \R{R1top_appendix}] gives us a
spatial condition for solution\R{R1top_appendix} to be valid: we must have $x \geq x_{dh}(t)$, where

\[
x_{dh}(t) \equiv L + \frac{\g}{2 \eta \vuh} d^{2} -
    \frac{1}{2} \g \eta \vuh (t + \frac{d}{\eta \vuh})^{2}.
\]

This is the mathematical origin of the `downhill front' that we
described intuitively in the main
 text as the limit of extension of the boundary effects originating in the upper edge of the cell.

\subsection{Central region}
 The evolution equation for $R$ is the same as
in the top zone, so that the differential equations giving the
characteristics also are the same [Eqs.\R{eqparam}]. But now we
must impose the initial condition $R(x,t=0)=0$, which gives the
following set of initial conditions for the characteristics:
$t(s=0)=0, \; x(s=0)= \xi,\; R(s=0)=0$ (and $0 \leq \xi \leq L$). We obtain:

\begin{eqnarray}
\label{tparam1central}
t(s) & = & s \\
\label{xparam1central}
x(s) & = & - \frac{1}{2} \g \eta \vuh \: s^{2} -\g d \: s + L \\
\label{Rparam1central}
R(s) & = & \eta \vuh \: s.
\end{eqnarray}

Combining \R{tparam1central} and \R{Rparam1central} gives an
explicit solution for $R$: $R(x,t)  = \eta \vuh t$.

This solution is valid in a certain spatial domain. It is limited
upwards by the top region [i.e.
 $x \leq x_{dh}(t)$]. It is also limited downwards by $x_{uh}(t)$, because at this point the form of
the evolution equation of $R$ changes (the amplification term
vanishes), and consequently does the form of the differential
equations that give the characteristics.

\subsection{Bottom region}

In this region, Eqs.\R{eqcarac} are given by:
$dt/ds = 1,\; dx/ds = - \g (R+d),\; dR/ds= 0$.

Here, the boundary condition is given by the continuity of $R$ at
the border of the central and the bottom zones: $R(x=x_{uh}(t),t)= \eta \vuh t$ for $t \geq 0$.
 This gives the initial conditions: $t(s=0)= \xi,\; x(s=0) = \vuh \xi,\; R(s=0)= \eta \vuh \xi$.
Solving and rewriting $R$ explicitly in terms of $x$ and $t$ leads
to the solution:

\begin{eqnarray}
\lefteqn{R(x,t) = -\; \frac{1}{2} \left(d + \frac{\vuh}{\g} - \eta \vuh t \right) }\nonumber
 \hspace{7mm} \\
 & & + \; \frac{\vuh}{2 \g} \sqrt{ \left(\frac{\g d}{\vuh} +1 - \g \eta\ t \right)^{2}
   + 4 \frac{\g \eta}{\vuh} (x + \g d t) }, \nonumber
\end{eqnarray}

valid for $0 \leq x \leq x_{uh}(t)$.



\end{multicols}

\newpage


\begin{figure}

\caption{The basic assumption of the BCRE picture is that there
 is a sharp distinction between immobile grains with a profile $h(x,t)$,
 and rolling particles with a local amount $R(x,t)$. The immobile grains
 constitute the `static phase' and the rolling ones the `rolling phase'.
 The local slope of the static profile is called $\th(x,t)$.
}
\label{BcrePicture}

\end{figure}


\begin{figure}

\caption{Example of an open cell, so as to let the rolling material flow out.
 We suppose that the avalanche starts precisely at $\theta = \tm$
  (see text).
}
\label{figOpenCell}

\end{figure}


\begin{figure}

\caption{The profile of the static grains for $0 < t < t_{2}$. At the left of $x_{uh}(t)$,
 the slope has relaxed to its final value $\tn$. On the right, it still has the initial
 angle $\tm$.
}
\label{figh}

\end{figure}


\begin{figure}

\caption{ {\it (a)} Position (dotted lines) and motion (arrows) of the `downhill' and
`uphill' fronts during Stage I. The respective sizes of the static phase (dark) and
 rolling phase (light) have been modified for clarity purposes.
The positions of the fronts naturally define three regions, with specific physical meaning (see text):
bottom (1), central (2), and top (3). {\it (b)} Evolution of the rolling phase in the cell during Stage I.
 The plot presents $R$ vs.
the position $x$, at successive dates ($R$ and $x$ are given in grain diameters
$d$, and the parameters are
 $\vuh = 3 \g d$, $\eta = 0.1$ rad and $L = 1000 d$. Note the different horizontal and vertical
 scales). The amount of rolling grains grows with time in the whole cell. Regions borders correspond
to slope discontinuities. For $t=t_{1}$, the
profile presents a cusp, where the maximum thickness $R_{max}$ of the avalanche is reached.
 }
\label{figStageI}

\end{figure}


\begin{figure}

\caption{{\it (a)} `Downhill' and `uphill' fronts during Stage II. The fronts naturally define three
 spatial regions: bottom (1), central (2), and top (3). {\it (b)} Evolution of the rolling phase in
the cell during Stage II. The plot presents $R$ vs.
the position $x$ at successive dates from $t=t_{exit}$ to $t=t_{2}$ ($R$ and $x$ are
given in grain diameters $d$; $\vuh$, $\eta$
 and $L$ as in previous figures. Note the different horizontal and vertical
 scales). The amount of rolling grains globally decreases in
the cell. Regions borders correspond
to slope discontinuities. At $t=t_{exit}$, the peak amplitude $R_{max}$ reaches the bottom
end of the cell.}
\label{figStageII}

\end{figure}


\begin{figure}

\caption{{\it (a)} Both fronts have reached the cell borders (in most cases; see text),
there is only one region in the cell. {\it (b)} Evolution of the rolling phase in the cell during Stage III
($R$ and $x$ are given in grain diameters $d$; $\vuh$, $\eta$ and $L$ as in previous figures).
The profile at the beginning of the stage ($t=t_{2}$)
is progressively convected
 downhill, but dilates at the same time, as seen on the plots at $t=t_{2} + 0.05(t_{end} - t_{2})$
 and $t=t_{2} + 0.2(t_{end} - t_{2})$. This is because thicker vertical slices roll faster
 than thinner ones. We see that the grains at the top edge of the cell
 ($x = L$) at time $t=t_{2}$ are the last ones to evacuate; the avalanche is extinct uphill ($R=0$).
}
\label{figStageIII}

\end{figure}


\begin{figure}

\caption{Loss of material at the bottom edge of the cell as a function of
time. The loss is given by the flow rate $Q(x=0,t)$
($Q$  is in units of $\g d^{2}$, $t$ in units of $\g^{-1} \sim (d/g)^{1/2}$;
 $\vuh$, $\eta$ and $L$ as in previous figures).
Solid line: predicted shape with a linear velocity profile in the
flow; dashed line: predicted shape in the case of a constant velocity profile in the
flow, from Boutreux {\it et al.}, with the choice $v=\vuh$ (see text).
}
\label{figRoutput}

\end{figure}


\begin{references}



\bibitem{Nedderman}
For a detailed description of the Coulomb approach and its
extensions, see R.M. Nedderman,
 {\it Statics and Kinematics of Granular Materials} (Cambridge University
 Press, Cambridge, 1992).
\bibitem{RajchenbachCargese}
J. Rajchenbach, in {\it Physics of Dry Granular Media}, H.J.
Hermann, J.P. Hovi,
 and S. Luding eds.\ , 421 (Kluwer Academic Publishers, Dordrecht,1998).
\bibitem{PGGJapon}
P.-G. de Gennes, in {\it From Rice to Snow}, Lecture at the Nishina
Memorial Foundation, Publication Nr 38 (Nishina Memorial
Foundation, April 1998). See also Ref.\C{Boutreux98}.
\bibitem{Radjai1}
F. Radjai, D. Wolf, and S. Roux, {\it Phys. Rev. Lett.} {\bf 77},
274 (1996).
\bibitem{Radjai2}
F. Radjai, D. Wolf, S. Roux, M. Jean, and J.-J. Moreau, in {\it
Powders and Grains 97}, R. Behringer
 and J.Jenkins eds.\ , 455 (Balkema Publishers, Rotterdam, 1997).
\bibitem{Liu}
C.H. Liu, S. Nagel, D. Shechter, S. Coppersmith, S. Majumdar, O.
Narayan, and T. Witten,
 {\it Science} {\bf 269}, 513 (1995).
\bibitem{Bouchaud94}
J.-P. Bouchaud, M. E. Cates, J. Ravi Prakash, and S.F. Edwards,
{\it J. Phys. France I} {\bf 4}, 1383 (1994); J.-P. Bouchaud, M. E.
Cates, J. Ravi Prakash, and S.F. Edwards, {\it Phys. Rev. Lett.}
{\bf 74}, 1982 (1995). See also A. Mehta, in {\it Granular Matter},
A. Mehta ed.\ (Springer Verlag, Heidelberg, 1994).
\bibitem{note1}
In this article, we restrict ourselves to two-dimensional sandpiles.
\bibitem{PGGCras95}
P.-G. de Gennes, {\it C. R. Acad. Sci. Paris} {\bf 321-IIb}, 501
(1995).
\bibitem{PGGCoursCdF97}
P.-G. de Gennes, {\it Cours du Coll\`{e}ge de France}, Paris,
unpublished, 1997.
\bibitem{Boutreux98}
T. Boutreux, E. Rapha\"el and P.-G. de Gennes, {\it Phys. Rev. E}
{\bf 58}, 4692 (1998).
\bibitem{note2}
In Ref.~[\onlinecite{Boutreux98}], it was implicitly assumed that
the saturation length $\xi'$ was identical to $\xi$. In the present
article, we do not make this assumption. For physical reasons, we
expect $\xi'$ to be somewhat smaller than $\xi$ ($\xi' \sim 3 d$
and $\xi \sim 10 d$).
\bibitem{StopFlow}
T. Boutreux and E. Rapha\"el, {\it Phys. Rev. E} {\bf 58}, 7645
(1998).
\bibitem{BouchaudCargese}
J.P. Bouchaud and M.E. Cates, in {\it Physics of Dry Granular
Media}, H.J. Hermann, J.P. Hovi,
 and S. Luding eds.\ , 465 (Kluwer Academic Publishers, Dordrecht,1998).
\bibitem{note6}
More precisely, $v_{0}$ and $\g$ should depend on the slope of the
sandpile. Since, in the situation we are going to consider in this
paper, the slope will never depart from $\tm$ by more than a few
degrees, we might write to first order $v_{0} \sim (gd \sin(\tm))^{1/2}$
and $\g \sim (g \sin(\tm)/d)^{1/2}$; see for
instance\C{RajchenbachCargese}.
\bibitem{Douady}
 S. Douady, B. Andreotti, and A. Daerr, submitted to {\it Eur. Phys. J. B}.
\bibitem{DuranMRS}
J. Rajchenbach, E. Cl\'ement and J. Duran, in {\it Fractal Aspects
of  Materials}, F. Family,
 P. Meakin, B. Sapoval and R. Wool eds.\, Proccedings of the M.R.S., {\bf 367}, 525 (1995).
\bibitem{Azanza}
E. Azanza, P. Chevoir and P. Moucheron, in {\it Powders and Grains
97}, R. Behringer and J. Jenkins,
 eds.\ , 455 (Balkema Publishers, Rotterdam, 1997).
\bibitem{Pouliquen}
O. Pouliquen, to appear in {\it Phys. Fluid}.
\bibitem{note3}
Boutreux {\it et al.} have shown in Ref.\C{Boutreux98} that the
notion of `repose angle' is not an intrinsic property of the
material, but is explicitly dependent on the cell configuration.
\bibitem{characteristics}
See for example: D. Zwillinger, {\it Handbook of Differential
Equations}, 2nd edition, p. 368
 (Academic Press, New York, 1992), or P. R. Garabedian, {\it Partial Differential Equations},
 p.18 (Wiley \& Sons, New York, 1964).
\bibitem{Bronshtein}
I. Bronshtein and K. Semendyayev, {\it Handbook of Mathematics},
3rd edition (Springer Verlag, Heidelberg, 1998).
\bibitem{note4}
In the following expression, the determination of the cubic root
shall be chosen in the following
 way: let $z=re^{i \theta}$ be a complex number ($r \geq 0, -\pi < \theta \leq \pi$), then
                        $\sqrt[3]{z}=\sqrt[3]{r} \; e^{i \frac{\theta}{3}}$.
\bibitem{note5}
In situations where $t_{exit} \ll t_{2}$ is not fulfilled, the
description of the avalanche is slightly modified: the bottom
region still exists at the beginning of Stage III. Thus the
$R$-profile left at the end of Stage II is given by
Eq.\R{R2central} for $x \geq x_{dh}(t=t_{2})$, and by
 Eq.\R{R1bottom} for $x \leq x_{dh}(t=t_{2})$. Then, this profile is convected in the same way as
described in the main text for the case $t_{exit} \ll t_{2}$.
\bibitem{BoutreuxdeGennes96}
T. Boutreux and P.-G. de Gennes, {\it J. Phys. France I} {\bf 6},
1295 (1996).
\bibitem{TheorieMakse}
H. A. Makse, {\it Phys. Rev. E} {\bf 56}, 7008 (1998); H. A. Makse,
P. Cizeau and H. E. Stanley, {\it Phys. Rev. Lett.} {\bf 78}, 3298
(1997).

\end{references}
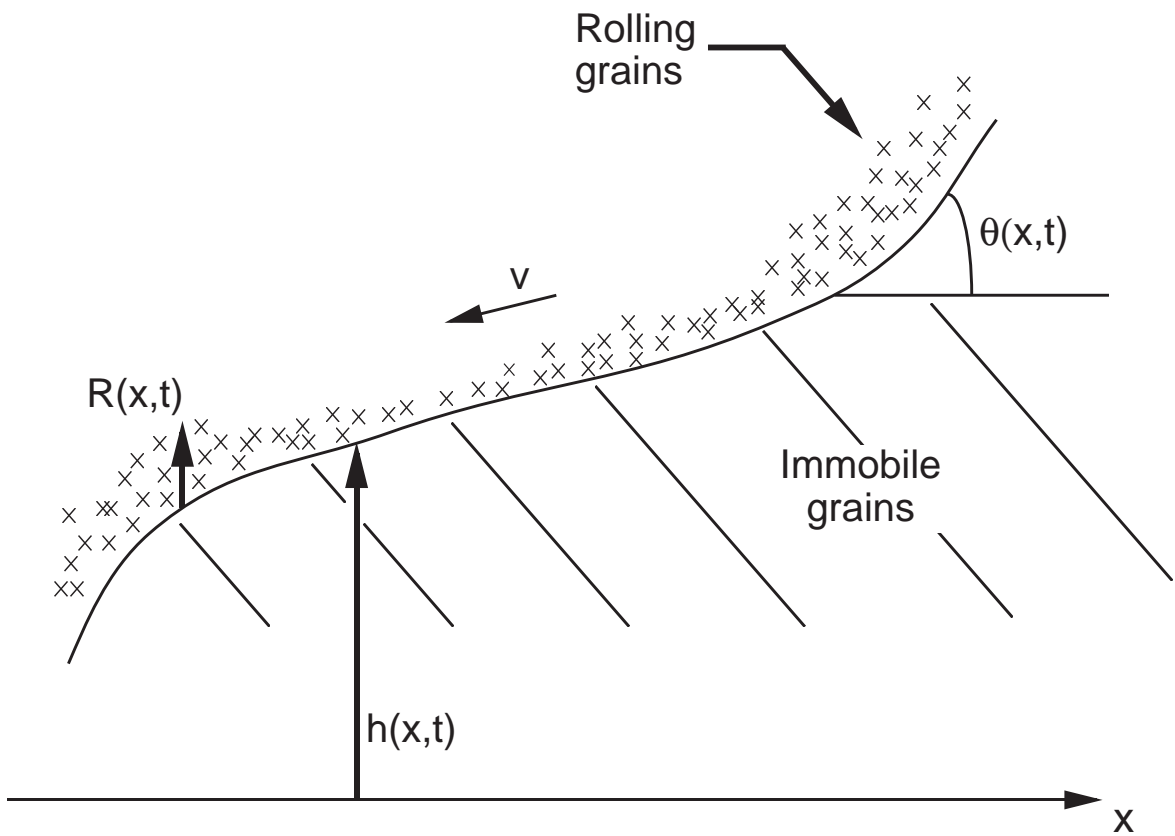
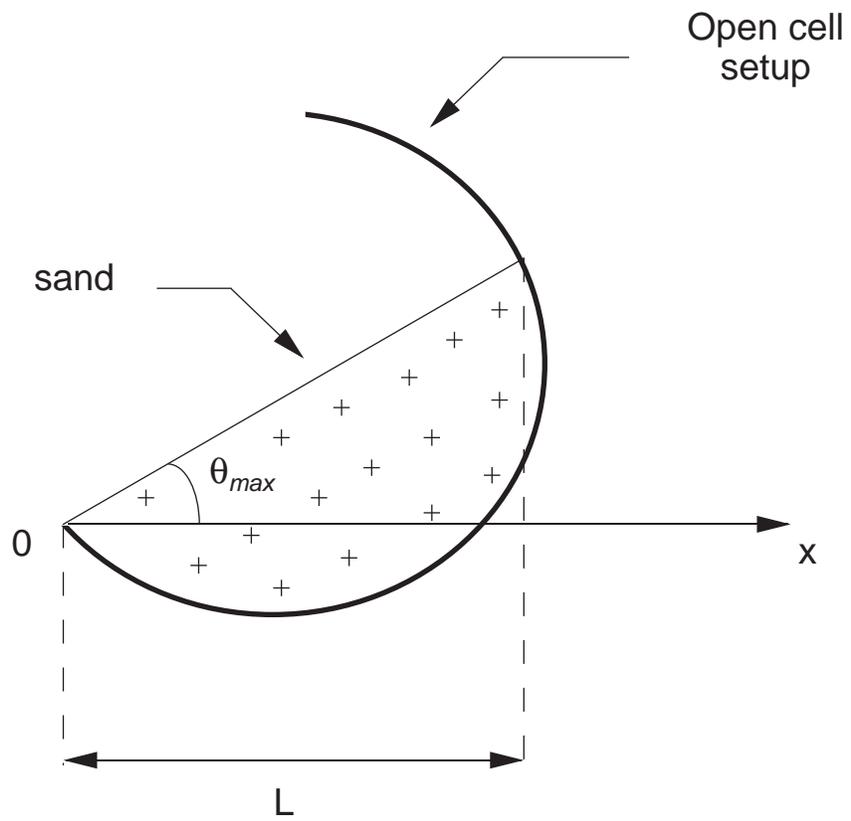
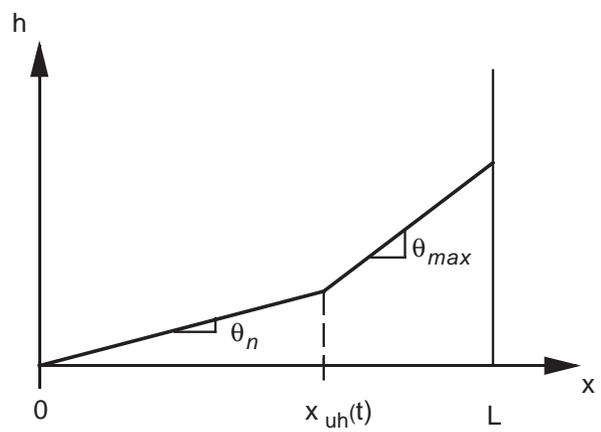
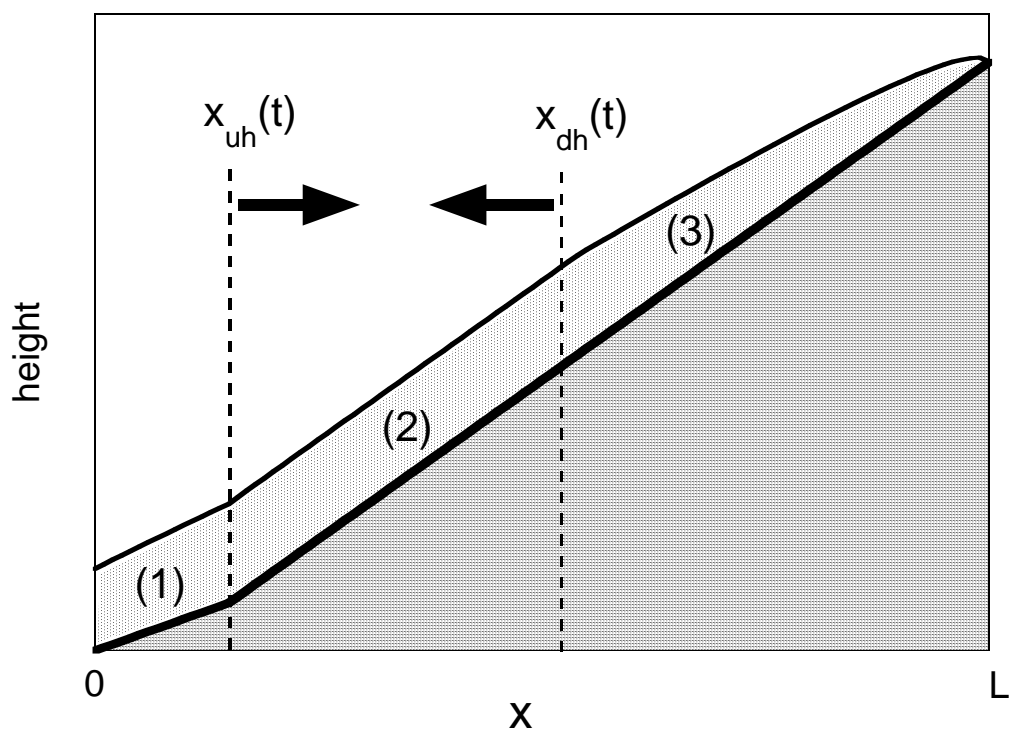
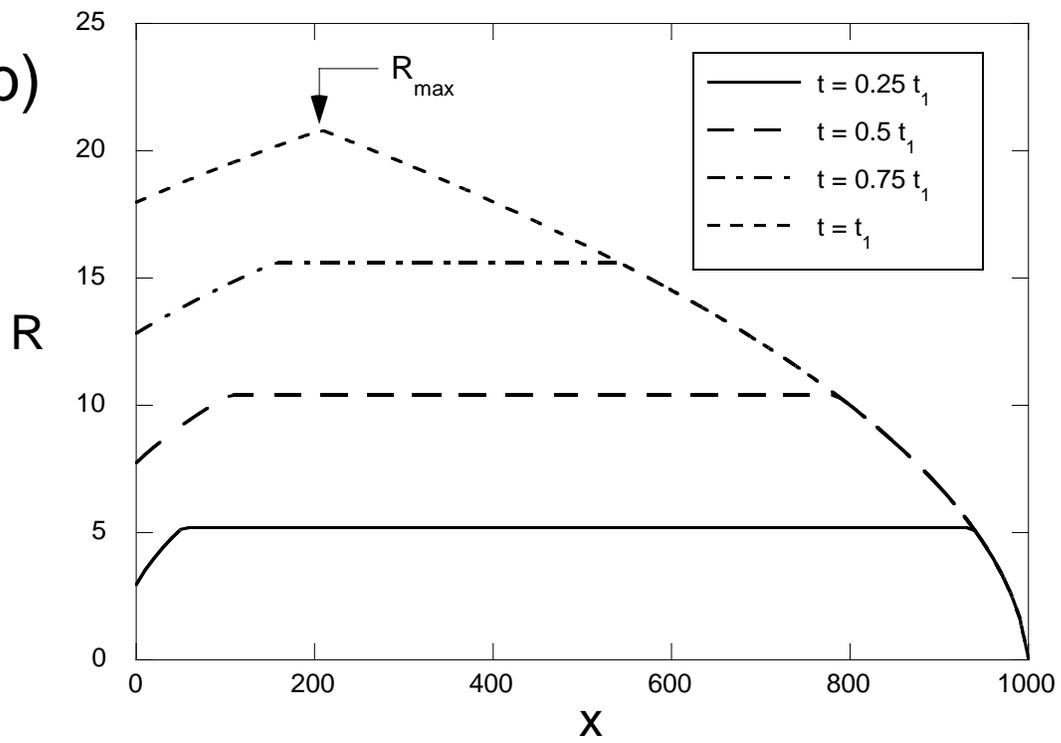
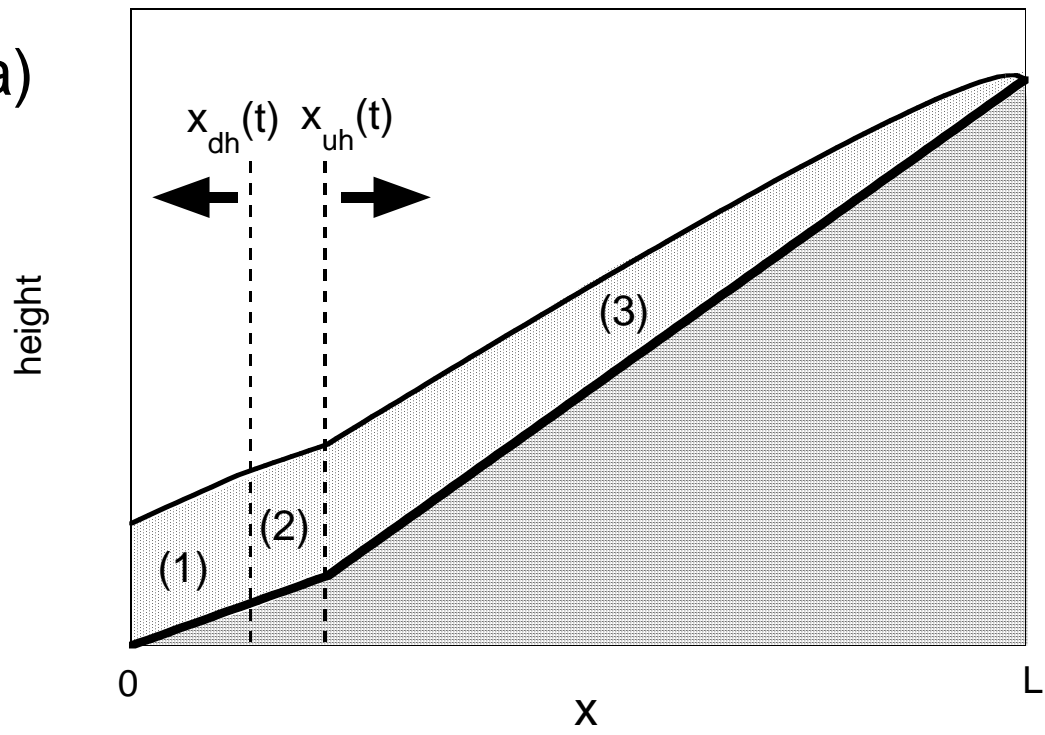
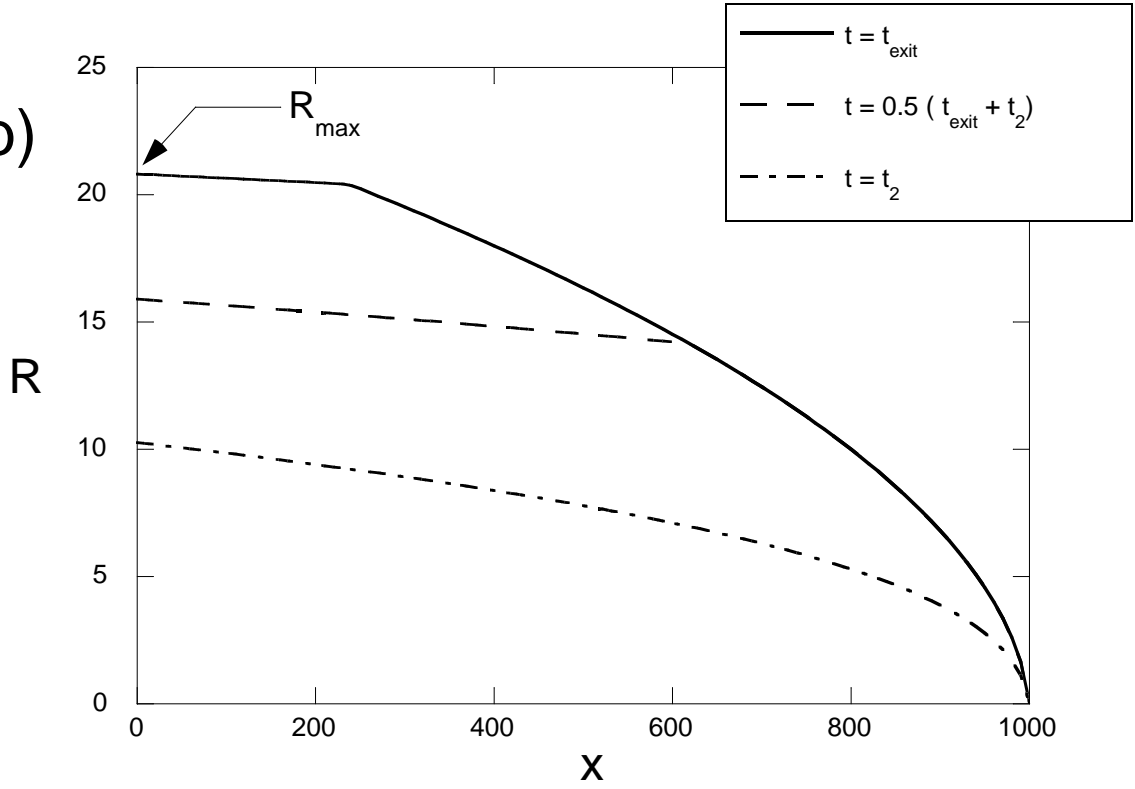
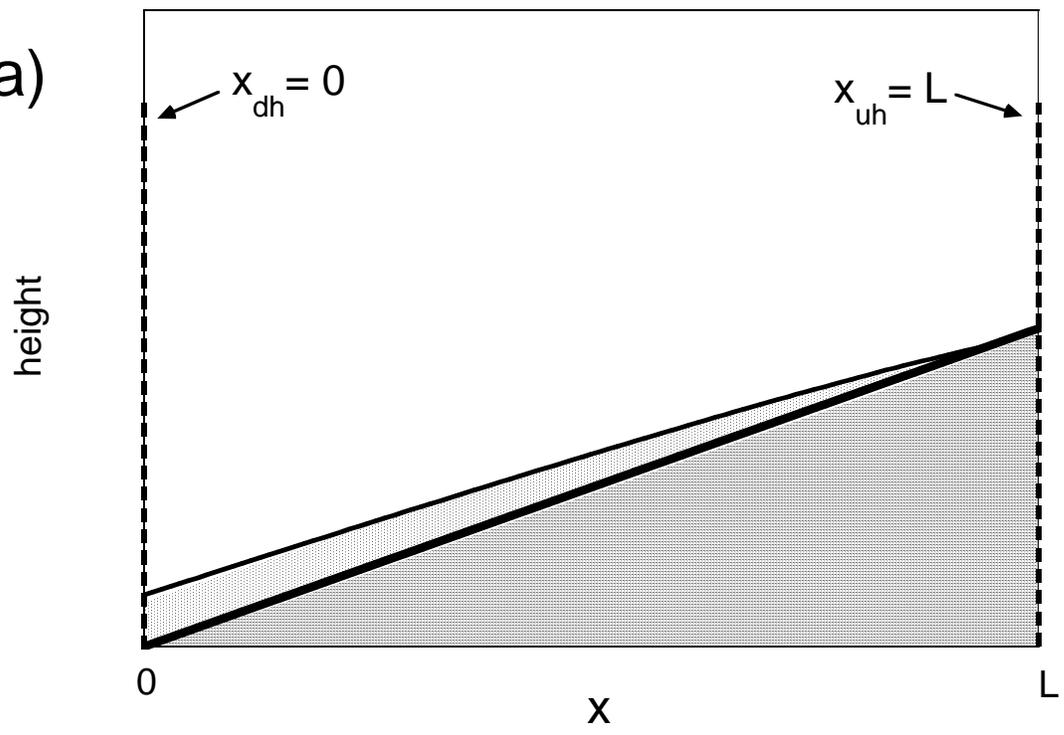
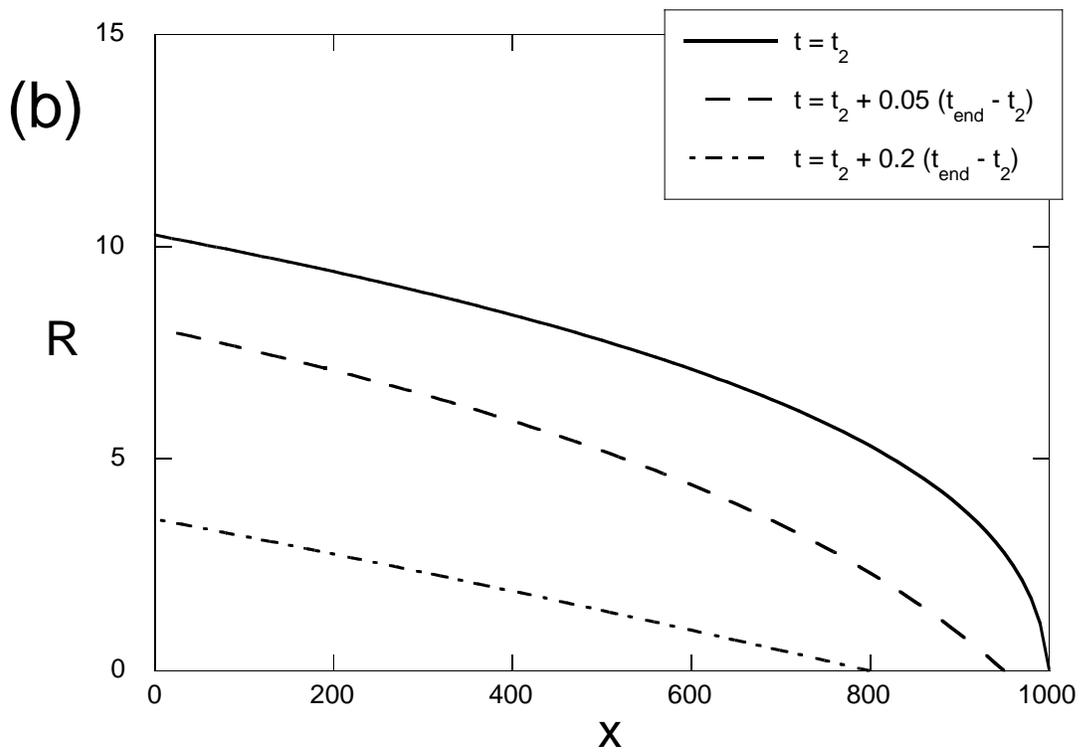
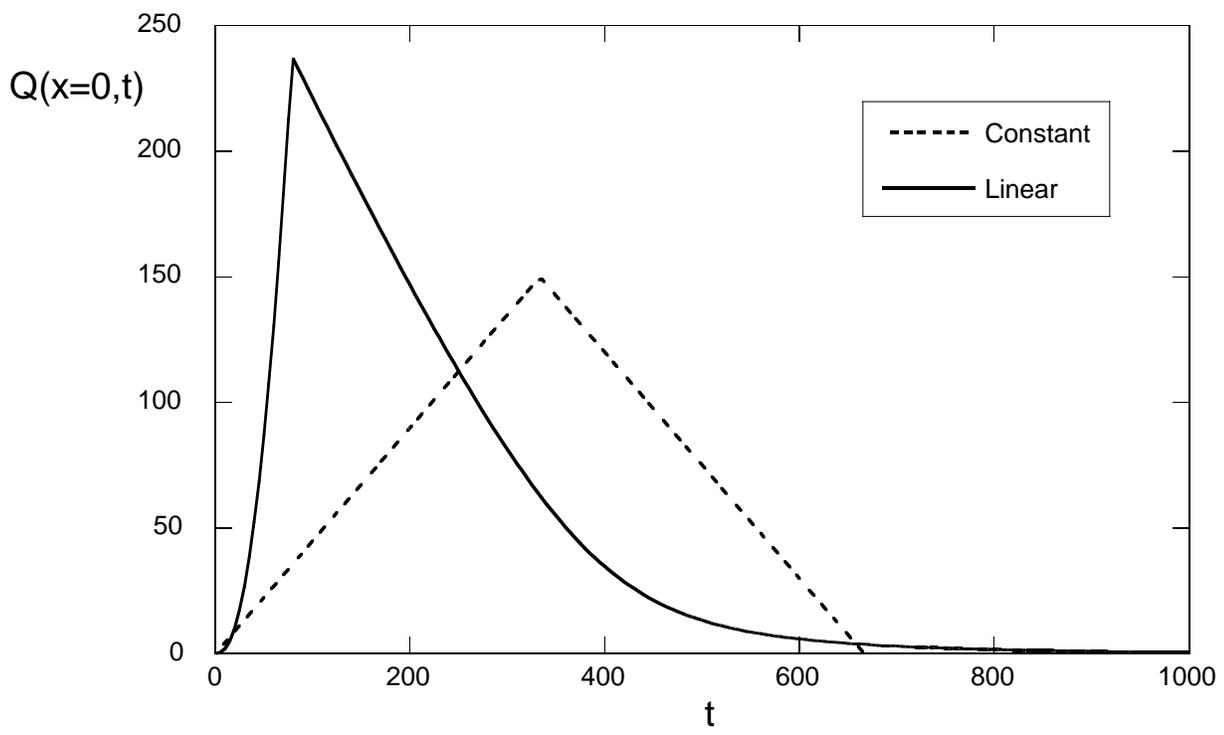
\end{document}